\documentclass[12pt]{article}
\usepackage{amsfonts,amsmath,amsthm,amssymb,times}
\usepackage{graphicx}
\usepackage{setspace}
\usepackage{amsmath}
\usepackage{upgreek}
\usepackage{enumerate}
\usepackage{color}
\usepackage{amsfonts}
\usepackage{amssymb}%
\usepackage{rotating}
\usepackage{subfig}
\usepackage{verbatim}
\usepackage{hyperref}
\usepackage{float}

\newtheorem{definition}{DEFINITION}[section]

\begin{document}
\title{ \bf Testing for the Equality of two Distributions on High Dimensional Object Spaces}

\author{ R. Guo$\thanks{Research
supported by
National Science Foundation Grant DMS-1106935}$, V. Patrangenaru $\thanks{Research
supported by National Science Foundation Grant DMS-1106935 and the
National Security Agency Grant MSP-
H98230-15-1-0135}$
\\ Florida State University, USA
}
\maketitle
\doublespacing
\section{Introduction}
Assume $X_1,X_2,Y_1,Y_2$ are independent random vectors $X_1,X_2,Y_1,Y_2$ in $\mathbb{R}^d,$ with  $\mathbb P_{X_1} =\mathbb P_{X_2}= Q_1, \mathbb P_{Y_1} = \mathbb P_{Y_2}=Q_2.$ Sz\'ekely and Rizzo (2013) defined a nonnegative parameter, called {\em energy distance }, that is associated with the pair of distributions $Q_1, Q_2,$ is labelled here $\varepsilon(Q_1, Q_2),$ and is defined as follows
\begin{equation}\label{energy-distance}
\varepsilon(Q_1, Q_2) = 2E\|X_1-Y_1\|-E\|X_1-X_2\|-E\|Y_1-Y_2\|,
\end{equation}
where $\|\cdot\|$ is the Euclidean norm.
The sample counterpart of the energy distance, associated with the independent random samples $X_1,\ldots,X_{n_1}$ from $Q_1$ and $Y_1,\ldots,Y_{n_2}$ from $Q_2,$ is called {\em energy statistic } and given by
\begin{eqnarray}\label{eq:en-manif}
\mathcal{E}_{n_1,n_2}=\frac{2}{n_1n_2}\sum_{i=1}^{n_1}\sum_{m=1}^{n_2}\|X_i-Y_m\|-\nonumber \\
\frac{1}{n_1^2}\sum_{i=1}^{n_1}\sum_{j=1}^{n_1}\|X_i-X_j\|-\frac{1}{n_2^2}\sum_{l=1}^{n_2}.
\sum_{m=1}^{n_2}\|Y_l-Y_m\|.
\end{eqnarray}
 In Section 2 we briefly review a nonparametric test for homogeneity ( $H_0: Q_1 = Q_2$ vs. $H_1: Q_1 \ne Q_2$) based on energy statistics that is . due to Sz\'ekely and Rizzo (2004). This test is based on the limiting distribution of the test statistic  $T_{n_1,n_2}=\frac{n_1n_2}{n}\mathcal{E}_{n_1,n_2},$ where the total sample size $n = n_1 + n_2$ goes to infinity, while, at the same time $\lim_{n\to\infty} n_1/n = \lambda \in (0,1).$ According to Sz\'ekely and Rizzo (2004), homogeneity is rejected for large values of $T_{n_1,n_2}.$ The significance of the energy test statistic is assessed using a permutation test (see also Wei et a.(2016)).

In Section 3 we introduce the extrinsic energy distance associated with two probability measures on an embedded object space. Assuming the {\em object space} $\mathcal M$  admitting a manifold stratification (see Patrangenaru and Ellingson (2015, p.475)) is embedded in an Euclidean space via a map $j: \mathcal M \to \mathbb R^N,$ we then adapt the results in Section 3 due to Sz\'ekely and Rizzo (2013) to test for homogeneity of distributions of two random objects (r.o.'s) on $\mathcal M$  by means of a test based on the extrinsic energy statistic associated with object data from such distributions. This test is validated via a simulation example, where we test for differentiating between two distributions of Kendall shapes of  planar $k$-ads, using a VW energy statistic based test.

Section 4 is dedicated to an application to medical imaging, where we test for the homogeneity for the distributions of Kendall shapes of the midsection of the Corpus Callosum (CC) in a clinically normal population and a population of individuals diagnosed with ADHD.
In principle the CC midsections data lies on the Hilbert manifold of direct similarity shapes of contours (see Ellingson et al.(2013), Patrangenaru and Ellingson (2015, Chapter 3)), however since the data available is discretized to fifty pseudo-landmarks, we conduct the test of homogeneity on a high dimensional Kendall shape space, with the VW-embedding into an Euclidean space of self adjoint complex matrices, and find out that the two distributions of CC midsections shapes are not significantly different, while there VW means are, due to a high dimensionality effect noticed by Bai and Saranadasa(1996).

\section{Energy statistics}
Energy statistics are estimators of the energy distance that depend on the distances between observations. The idea behind energy statistics is to consider a statistical potential energy that would parallel Newton's gravitational potential energy. This statistical potential energy is zero if and only if a certain null hypothesis relating two distributions holds true. In this paper we consider energy statistics based on distances, that are associated with a random sample $X_1,\ldots,X_n$ from a $d$-dimensional probability measure $Q$ and a kernel function $h:R^d\times R^d\rightarrow R,$
\begin{equation}\label{eq:V-stat}V_n=\frac{1}{n^2}\sum_{i=1}^n\sum_{j=1}^nh(X_i,X_j),
\end{equation}
were $h(X_i,X_j)=h(X_j,X_i)$ is a symmetric function in the Euclidean distances $\|X_i-X_j\|$ between sample elements.

The {\em energy statistic } associated with the independent random samples $X_1,\ldots,X_{n_1}$ from $Q_1$ and $Y_1,\ldots,Y_{n_2}$ from $Q_2,$ is given by
\begin{eqnarray}\label{eq:en-manif}
\mathcal{E}_{n_1,n_2}=\frac{2}{n_1n_2}\sum_{i=1}^{n_1}\sum_{m=1}^{n_2}\|X_i-Y_m\|-\nonumber \\
\frac{1}{n_1^2}\sum_{i=1}^{n_1}\sum_{j=1}^{n_1}\|X_i-X_j\|-\frac{1}{n_2^2}\sum_{l=1}^{n_2}
\sum_{m=1}^{n_2}\|Y_l-Y_m\|.
\end{eqnarray}
Let $n = n_1+n_2$ be the total sample size of the pooled sample, and let us call $T_{n_1,n_2}=\frac{n_1n_2}{n}\mathcal{E}_{n_1,n_2}$ {\em t-energy statistic}. The t-energy statistic can be used
for testing equality $Q_1 = Q_2.$ The hypothesis of equal distributions is rejected for large values of $T_{n_1,n_2}.$
Under null hypothesis, a random permutation of the pooled sample has the same distribution as a random sample size of $n$ from the mixture $W$, where $W$ is the random variable obtained by sampling from the distribution of $X_1$ with probability $n_1/n$ and from the distribution of $Y_1$ with probability $n_2/n$. Assume $\alpha\in (0,1)$ is fixed and let $c_\alpha$ be the constant such that
$$\lim_{n\rightarrow\infty}P(T_{n_1,n_2}>c_\alpha)=\alpha.$$

The proof of the existence of $c_\alpha$ is given in Sz\'ekely and Rizzo (2004), along with proofs on the limiting distribution of $T_{n_1,n_2}$ and its consistency. There is also shown that the null  hypothesis $H_0: Q_1 = Q_2$ is rejected at level $\alpha$ if $T_{n_1,n_2}>c_\alpha,$ and that the test is consistent against general class of alternatives, provided $n_1/n$ converges to a constant in $(0,1).$

\section{Extrinsic energy statistics for object data}
\subsection{General energy statistics}
Big data of high complexity observations of such as axes, shapes, trees, diffusion tensors can be regarded as points on certain metric spaces, known as object spaces ( see Patrangenaru and Ellingson (2015), Section 3.5.). Arguably, to date, all object spaces encountered in Statistics have a structure of metric spaces that admits a manifold stratification ( see Patrangenaru and Ellingson (2015), xvii). From now on, we will therefore assume that an object space is a stratified space (has a manifold stratification), unless otherwise specified. Given a stratified space, an embedding is a one to one homeomorphism on its image, that is an embedding when restricted to each manifold like stratum.

\begin{definition} The extrinsic energy distance between two independent probability distributions $Q_1$ and $Q_2$ on the $d$ dimensional object space $\mathcal M$ can therefore be considered relative to an embedding $j:\mathcal M \to \mathbb R^N,$ and is defined as follows
\begin{equation}\label{eq:energy-ob}\varepsilon_j(Q_1,Q_2)= 2E\|j(X)-j(Y)\|-E\|j(X)-j(X')\|-E\|j(Y)-j(Y')\|,
\end{equation}
assuming $E\|j(X)\|<\infty,E\|j(Y)\|<\infty$, $X, X',Y,Y'$ are independent r.o.'s with $\mathbb P_X = \mathbb P_{X'} = Q_1$ and $\mathbb P_Y = \mathbb P_{Y'} = Q_2.$
\end{definition}

 We can therefore consider the sample counterpart of the energy distance on an embedded object space, called {\em extrinsic energy statistic}, given by:
\begin{eqnarray}\label{eq:exen-linear}
\mathcal{E}_{j,n_1,n_2}(X,Y)=\frac{2}{n_1n_2}\sum_{i=1}^{n_1}\sum_{m=1}^{n_2}\|j(X_i)-j(Y_m)\|-\nonumber \\
\frac{1}{n_1^2}\sum_{i=1}^{n_1}\sum_{h=1}^{n_1}\|j(X_i)-j(X_h)\|-\frac{1}{n_2^2}\sum_{l=1}^{n_2}\sum_{m=1}^{n_2}\|j(Y_l)-j(Y_m)\|,
\end{eqnarray}
where $X_1,\ldots,X_{n_1}$, $Y_1,\ldots,Y_{n_2}$ are i.i.d.r.o.'s from $Q_1$ and $Q_2$ respectively.

From Section 2, if we set $T_{j, n_1,n_2}=\frac{n_1n_2}{n}\mathcal{E}_{n_1,n_2},$ it follows that given an embedding $j:\mathcal M \to \mathbb{R}^N,$ and a level of significance $\alpha \in (0,1),$ there is a $c_\alpha$ such that the null hypothesis $H_0: Q_1 = Q_2$ is significant at level $\alpha$ if ${T}_{j,n_1,n_2}>c_\alpha.$ The test is consistent against general class of alternatives provided $n_1/n$ converges to a constant in $(0,1).$ Under $H_0,$ the t-energy statistic ${T}_{j,n_1,n_2}$ depends on the unknown distribution $Q = Q_1 = Q_2$ via the selected pooled sample of size $n.$ We then use a nonparametric bootstrap methodology, to test whether should we reject the null hypothesis; however instead of using a permutation test ( see Sz\'ekely and Rizzo (2004)), we use a nonparametric bootstrap test based on sampling with replacement.

Consider $X$ and $Y$ are independent random objects, with $P_X=Q_1$ and $P_Y= Q_2$. Let $A=\{X_1,\ldots,X_{n_1}, Y_1,\ldots, Y_{n_2}\}$ be the pooled sample, $X_a, Y_b$ being i.i.d.r.o.'s from $Q$ under $H_0.$ Conditionally on $X_a, a = 1,\ldots, n_1$, respectively on $Y_b, a = 1,\ldots, n_2,$ we obtain the pooled sample $A^*$ of size $n,$ given by $A^* =  \{X_1^*,\ldots,X_{n_1}^*, Y_1^*,\ldots, Y_{n_2}^*\}.$

We obtain t-energy statistics from
\begin{eqnarray}\label{eq:exten}
{T}_{j,n_1,n_2}(X,Y)=\frac{n_1n_2}{n}\{\frac{2}{n_1n_2}\sum_{i=1}^{n_1}\sum_{m=1}^{n_2}\|j(X_i)-j(Y_m)\|-\nonumber \\
\frac{1}{n_1^2}\sum_{i=1}^{n_1}\sum_{h=1}^{n_1}\|j(X_i)-j(X_h)\|-\frac{1}{n_2^2}\sum_{l=1}^{n_2}\sum_{m=1}^{n_2}\|j(Y_l)-j(Y_m)\|\},
\end{eqnarray}
 and bootstrap t-energy statistics from
\begin{eqnarray}\label{eq:exbten}
{T}^*_{j,n_1,n_2}(X^*,Y^*)=\frac{n_1n_2}{n}\{\frac{2}{n_1n_2}\sum_{i=1}^{n_1}\sum_{m=1}^{n_2}\|j(X^*_i)-j(Y^*_m)\|-\nonumber \\
\frac{1}{n_1^2}\sum_{i=1}^{n_1}\sum_{h=1}^{n_1}\|j(X^*_i)-j(X^*_h)\|-\frac{1}{n_2^2}\sum_{l=1}^{n_2}\sum_{m=1}^{n_2}\|j(Y^*_l)-j(Y^*_m)\|\},
\end{eqnarray}

Under null hypothesis, since $Q_1$ and $Q_2$ are identically distributed, the $X_1^*,\ldots,X_{n_1}^*$ and $Y_1^*,\ldots, Y_{n_2}^*$ from bootstrap resampling should have same distribution as $X_1,\ldots,X_{n_1}$ and $Y_1,\ldots, Y_{n_2}$. Therefore, the t-energy statistic ${T}_{j,n_1,n_2}$ calculated from $A=\{X_1,\ldots,X_{n_1}, Y_1,\ldots, Y_{n_2}\}$ should follow same distribution as ${T}^*_{j,n_1,n_2}$ calculated from resampling observations $A^* =  \{X_1^*,\ldots,X_{n_1}^*, Y_1^*,\ldots, Y_{n_2}^*\}.$.

Let $\mathcal{T}^{(i)}_{j,n_1,n_2}$ be the notation of the i-th trial ${T}^*_{j,n_1,n_2}$ result from bootstrap resampling with replacement. And we estimate $P(\mathcal{T}_{j,n_1,n_2}<L)$ by $\frac{\#({T}^{(i)}_{j,n_1,n_2}<L)}{\#(Trials)}$ where $\#$ is the count function. Then we can estimate $c_\alpha$ from the definition of $P(\mathcal{T}_{j,n_1,n_2}<c_\alpha)=\alpha$ by reversed application of $P(\mathcal{T}_{j,n_1,n_2}<L)$. As the nonparametric test, we will reject the null hypothesis $Q_1=Q_2$ at confidence level $\alpha\%$ if $\mathcal{T}_{j,n_1,n_2}$ is larger than $(100-\alpha)\%$ of $\mathcal{T}^*_{j,n_1,n_2}$ from bootstrap resamplings.

\subsection{VW energy statistics on planar Kendall shape spaces}
\begin{figure}[H]%
    \centering
    \subfloat[Circle like landmarks]{{\includegraphics[width=6.2cm]{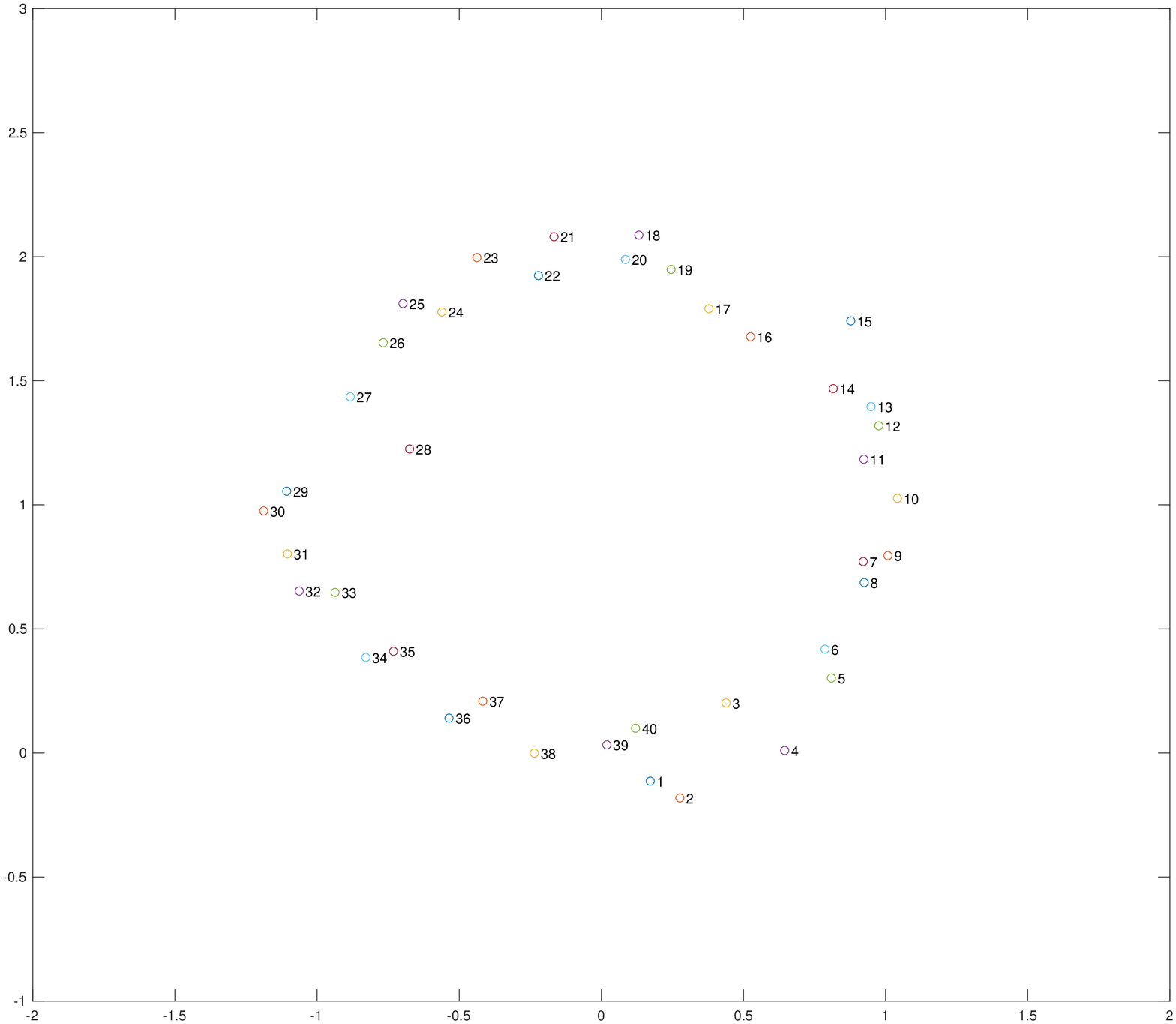} }}%
    \qquad
    \subfloat[Square like landmarks]{{\includegraphics[width=6.2cm]{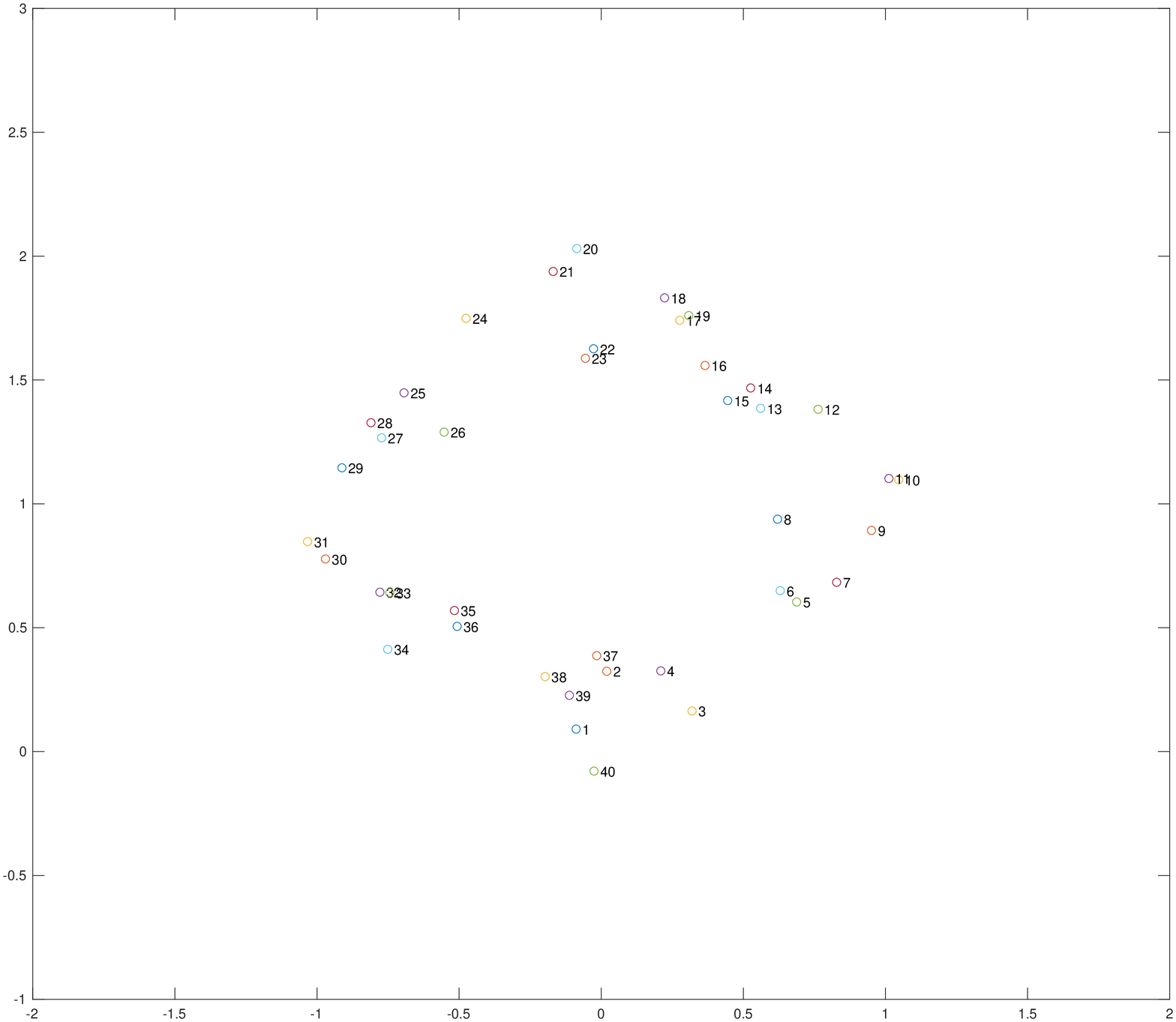} }}%
    \caption{\small Simulated $k$-ad from a contour like random configuration of forty labeled landmarks: round (left), square (right) }%
    \label{fig:example2}%
\end{figure}

A class of object spaces of great interest are those having a manifold structure. Since any $m$ dimensional manifold can be embedded in a numerical space (see Whitney (1944)), an extrinsic energy distance can be associated with any pair of probability on a manifold.
As an example, here we consider $\mathcal M = \mathbb CP^{k-2},$ which is the manifold representation of the planar Kendall shape space $\Sigma_2^k$ of 2D similarity shapes of $k$-ads (see Kendall(1984)). This manifold is usually embedded in the space of $(k-1)\times (k-1)$ complex selfadjoint  matrices via the Veronese-Whitney(VW) map $j$ given by
\begin{equation}
j([x])=xx^*,\|x\|=1,
\end{equation}
and the resulting chord distance is $\rho([x],[y]) = Tr((xx^*-yy^*)^2),$ assuming $\|x\|=1, \|y\|=1$ (see Kent (1992)). The corresponding energy distance between two distributions of Kendall shapes of k-ads will be called {\em VW-energy distance}, its sample counterpart is the {\em VW-energy statistic}, and the associated t-energy statistic is called {\em VW-t-energy statistic}. To test whether VW-t-energy statistic can distinguish between two distributions of Kendall shapes, we start with a simple simulation example. Beginning with two groups of forty landmark locations on a circular, respectively square contour, as shown in Figure \ref{fig:example} below,
\begin{figure}[H]%
    \centering
    \subfloat[Circular landmark configuration]{{\includegraphics[width=6.2cm]{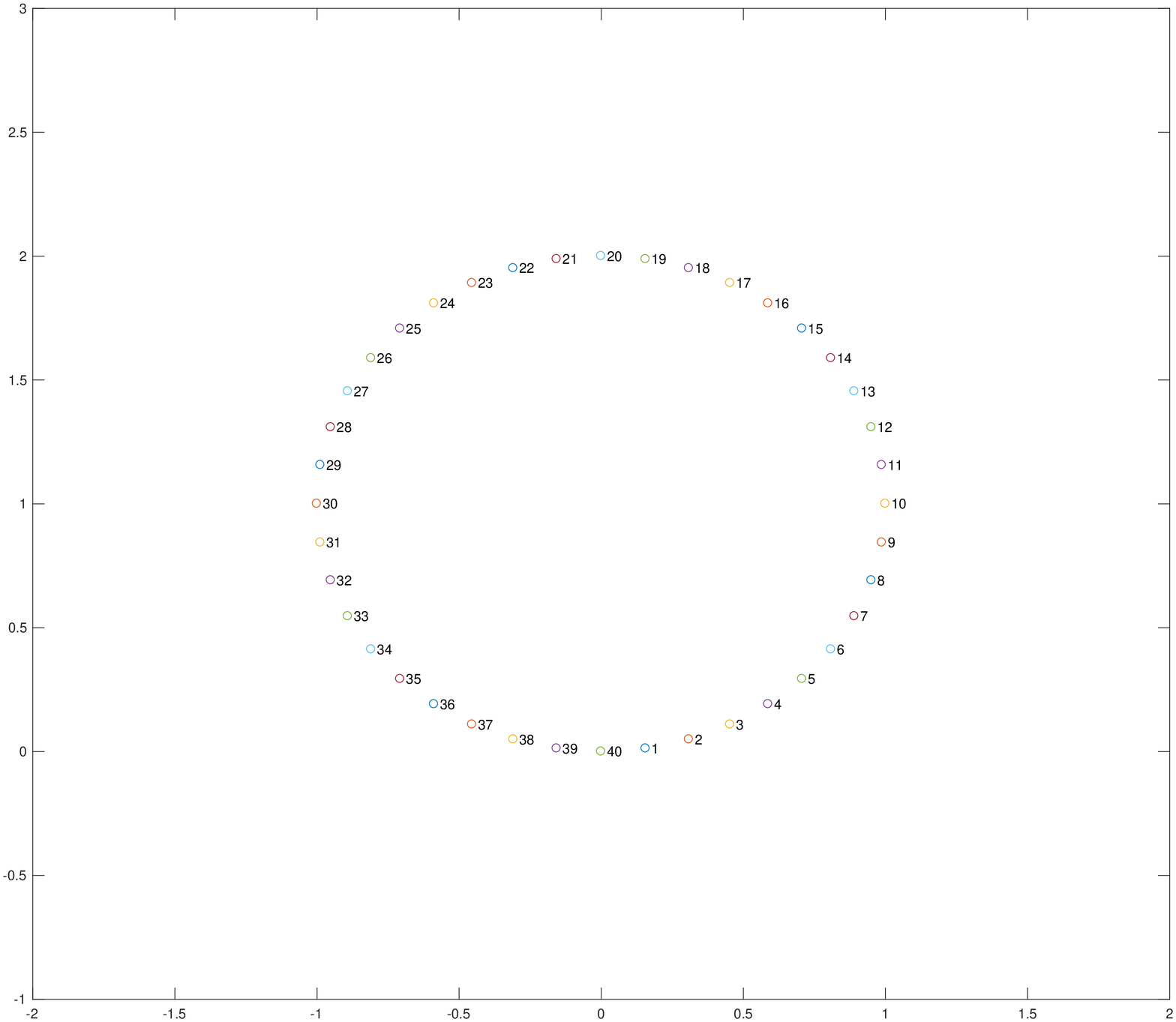} }}%
    \qquad
    \subfloat[Square landmark configuration]{{\includegraphics[width=6.2cm]{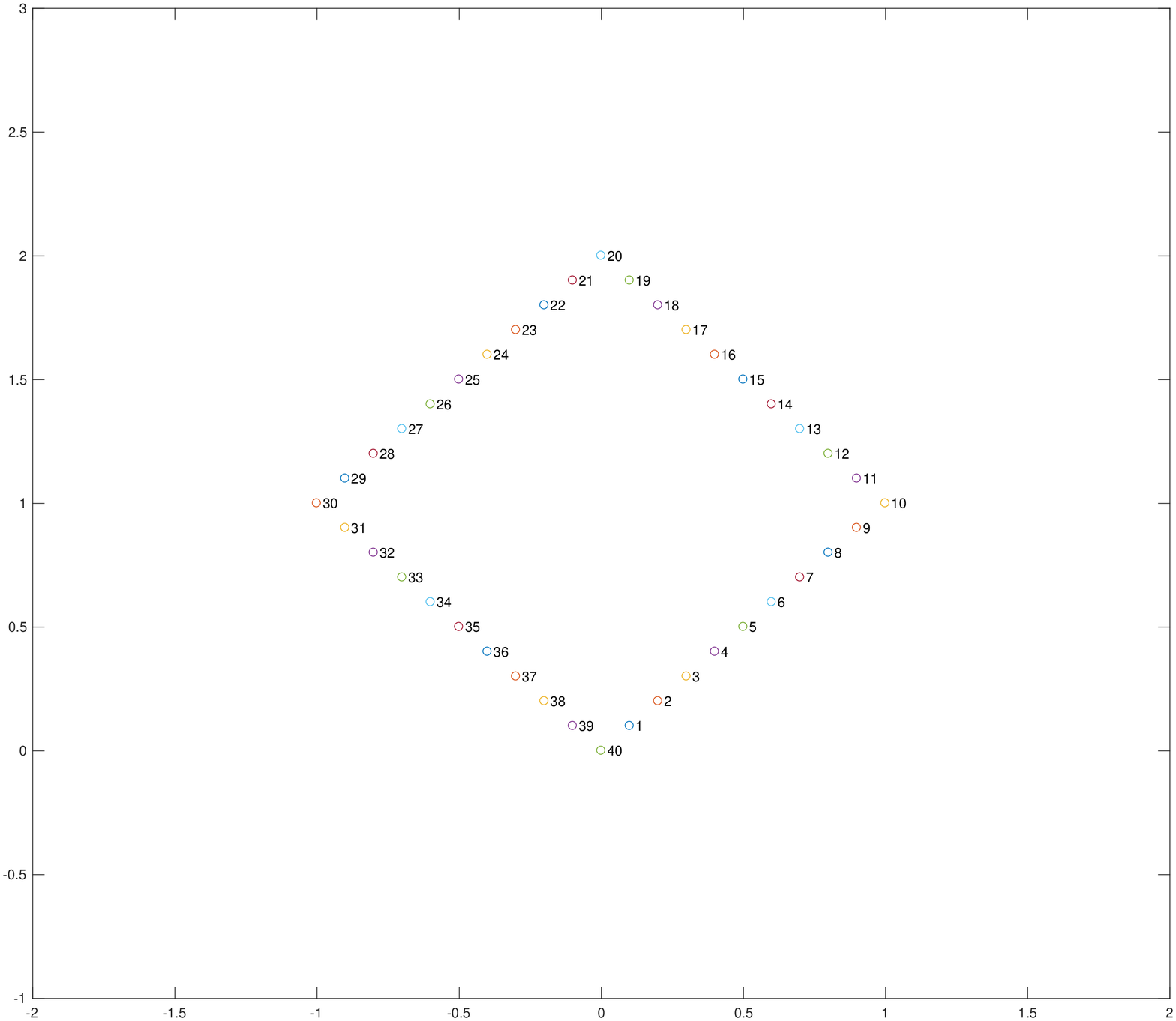} }}%
    \caption{\tiny Deterministic example of $k= 40$ labeled landmarks on two contours}%
    \label{fig:example}%
\end{figure}

we add random Gaussian white noise at each landmark location, and sample one hundred observations from the resulting two distributions of $k$-ads ( thus here $k=40, n_1 = n_2 = 100,$ as shown in Figure \ref{fig:example2}.

Let us label by $Q_r,$ respectively by $Q_s$ the distribution of Kendall shapes of $k$-ads obtained from the round contour, respectively from the square contour. Our null hypothesis is $H_0 : Q_r = Q_s.$

For the simulated data, the VW-t-energy statistic found was ${T}_{j,n_1,n_2}=3.5$. To test $H_0,$ we take a nonparametric bootstrap approach.
Under null hypothesis, there should be no difference between the two distributions of Kendall shapes, therefore the energy statistics should be the same no matter where the samples are chosen from: $Q_r$ or $Q_s$ because the random shapes are identically distributed. So we choose 1000 bootstrap resamples; we resample with replacement from the pooled pooled sample 100 random observations the first sample and again from second sample. We calculate the bootstrap VW-t-energy statistics {$T^*_{j,n_1,n_2}$} for these  two 1000 bootstrap resamples and and obtain the $c^*_\alpha$ from this 1000 resamples with replacement. Here $c^*_\alpha$ is defined as $\lim_{n\rightarrow\infty}P(T^*_{j,n_1,n_2}>c^*_\alpha)=\alpha,$ while $n_1/n$ is fixed.

For these 1000 resamples, we got $c^*_{0.05}=0.42, c^*_{0.01}=0.57$. Based on ${T}_{j,n_1,n_2}>c^*_{\alpha}, \alpha=0.01, 0.05$, it follows that the VW-energy test statistic shows that the two Kendall shape distributions are different at a high level of significance. So we can reject the null hypothesis that the distributions of round like shapes and squared like shapes are the same at level $\alpha = 0.01,$ thus confirming that these two populations of Kendall shapes are indeed generated from different contour shapes. This simulation result is an indication that the VW-energy distance based method can distinguish between two distributions of high dimensional Kendall shapes.

\begin{figure}[h]
\centering
\includegraphics[scale = 0.35]{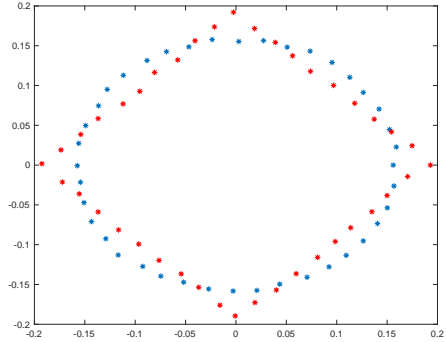}
\caption{\small The sample VW-means : round $k$-ad (blue), and square $k$-ad (red)}\label{vwsimulate}
\end{figure}

Note that the VW means of the two distributions of Kendall shapes are also different, as one can in Figure \ref{vwsimulate}, or even better, by using the two sample test form unmatched pairs of VW-means (see Patrangenaru and Ellingson (2015)).

\section{Corpus Callosum Shape and ADHD}

\begin{figure}[H]
\centering
\includegraphics[scale = 0.3]{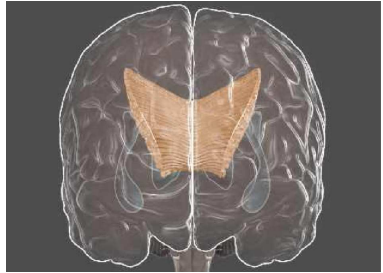}
\caption{\small 3D model of the CC}
\label{fig:3D}
\end{figure}

Attention-deficit/hyperactivity disorder (ADHD) is a brain condition marked by an ongoing pattern of inattention and/or hyperactivity-impulsive behavior that interferes with functioning or development in the young individual. ADHD is one of the most commonly diagnosed childhood behavioral disorders; arguably, it affects at least $5\%$ of school-age children (see Huang et al.(2015)). Some studies suggests that ADHD involves disfunction in wide functional networks of brain areas associated with attention and cognition and examine the structural integrity of white-matter neural pathways, which are the make-up of such functional networks, connecting fronto-striatal and fronto-parietal circuits in children with ADHD (see Silk et. al(2009)). Given that the largest mass of white matter in the brain is the Corpus Callosum (CC), here depicted in Figure \ref{fig:3D}, and 3D CC data is scarce, recently a study on the relationship between the CC shape and ADHD was focusing on the 2D direct similarity shape of the CC midsection (see Huang et al.(2015), Bhattacharya and Lin (2017)).

\begin{figure}[H]%
    \centering
    \subfloat[CC data - normal individuals ]{{\includegraphics[width=6.6cm]{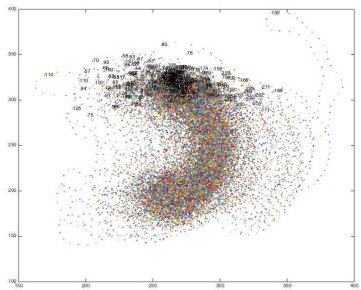} }}%
    \qquad
    \subfloat[CC data - ADHD diagnosed individuals ]{{\includegraphics[width=6cm]{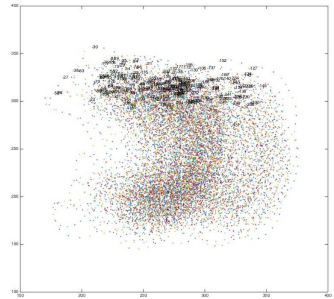} }}%
    \caption{\small ADHD-200 data of landmarks on contours of CC midsections }%
    \label{fig:obs}%
\end{figure}

\begin{figure}[H]%
    \centering
    \subfloat[One observation of $k$-ad - normal CC ]{{\includegraphics[width=6cm]{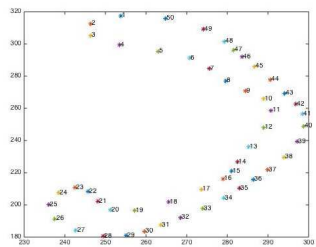} }}%
    \qquad
    \subfloat[One observation of $k$-ad -  ADHD CC ]{{\includegraphics[width=6cm]{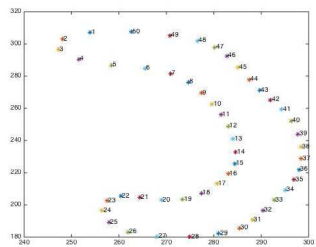} }}%
    \caption{\small Data point of $k=50$ labeled landmarks on the CC in normal individual (left ) vs and ADHD individual}%
    \label{fig:obs2}%
\end{figure}

Fifty landmarks were extracted from the midsections of the CC data from the
ADHD-200 Dataset - a data set that records both anatomical and resting-state functional MRI
data of 776 labeled subjects across 8 independent imaging sites, 491 of which were obtained from normally developing people and 285 in children and adolescents diagnosed with ADHD. In this section we are referring to this data set
as {\em CC data}. The CC data are coming originally from http://fcon$\_$1000.projects.nitrc.org/indi/adhd200/;
for convenience they are also posted at https://ani.stat.fsu.edu/$\sim$vic/ADHD-200CC.
In figure \ref{fig:obs} is displayed the unregistered CC data for each of the two groups separately. Note that, except for the differences due to sample sizes, the two distributions look somewhat similar. In addition, the landmark locations on the CC are given in Figure \ref{fig:obs2}

%\newpage

The VW-t-energy statistic for the two groups (normal vs ADHD diagnosed) is {$T_{j,n_1,n_2}=0.0948$. We calculate the bootstrap VW-t-energy statistics {$T^*_{j,n_1,n_2}$} for $N=500$ resamples, and obtain $c^*_{0.05}=0.1075, c^*_{0.1}=0.0958,$ showing that there no significant difference between the distributions $Q_{normal}$ and $Q_{ADHD}.$ Thus the shape of the CC-midsection does not seem to give a significant indication of the ADHD status of a child or adolescent.

\begin{figure}[H]%
    \centering
    \subfloat[ Shapes of normal CC data]{{\includegraphics[width=6cm]{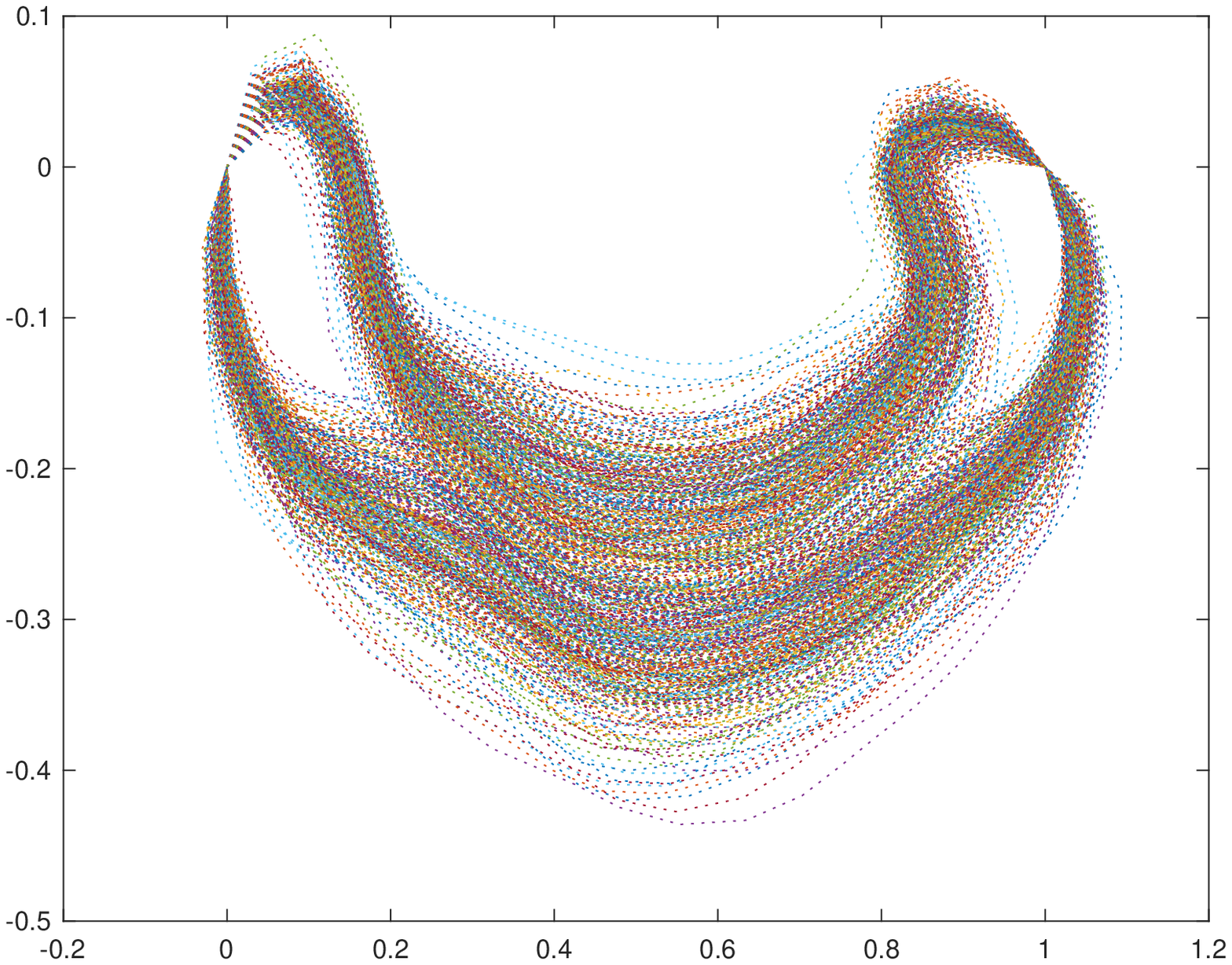} }}%
     \qquad
    \subfloat[ Shapes of ADHD CC data]{{\includegraphics[width=6cm]{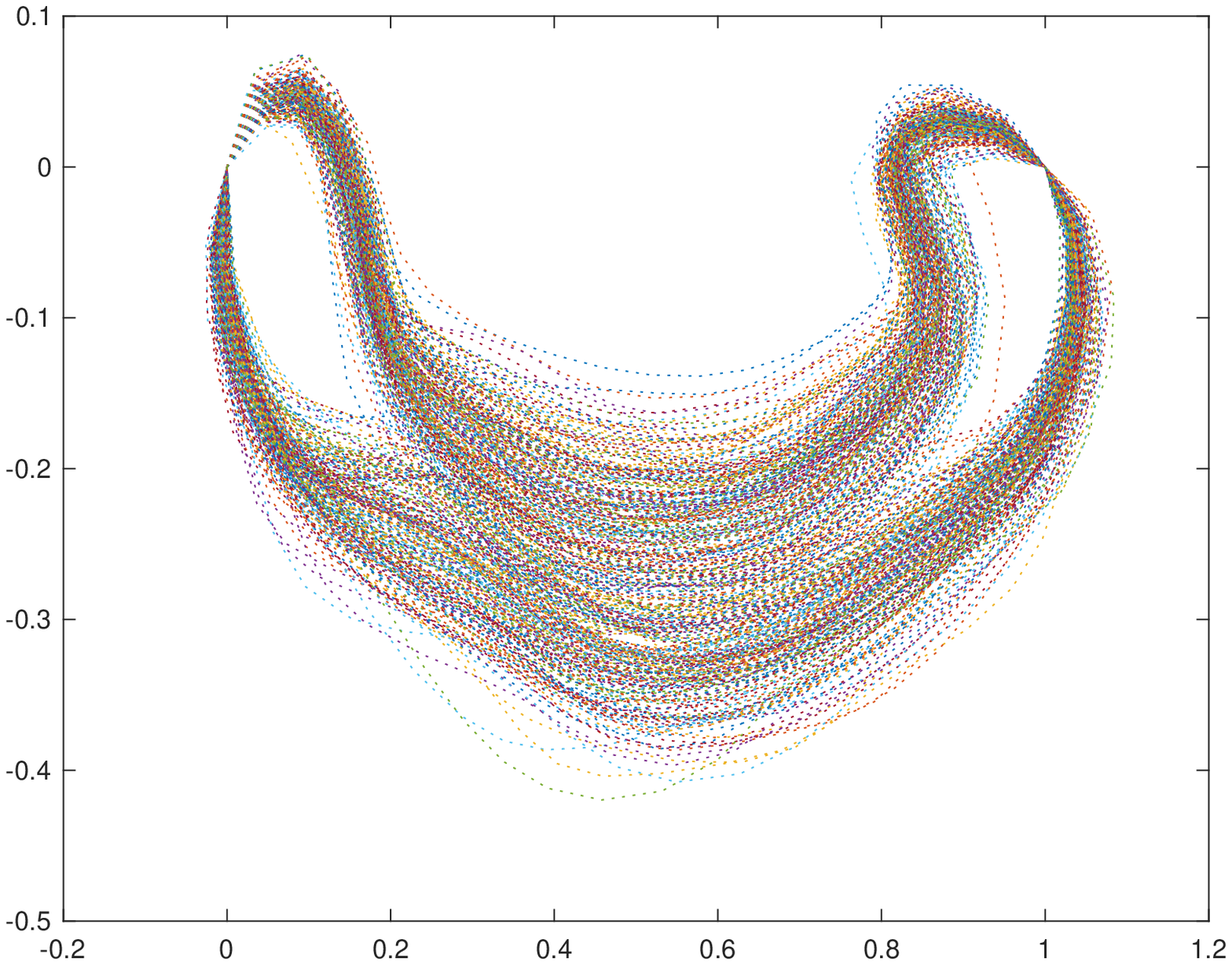} }}%
    \caption{\small Registered shapes of normal and ADHD CC midsection landmarks distributions}%
    \label{fig:example}%
\end{figure}

\section{Discussion}
The CC data was used in Bhattacharya et al. (2016, p. 312) as well as in Bhattacharya and Lin (2017) to test for the equality of the VW mean shapes of the normal and ADHD populations; the p value there is of the order $10^{-11}$, and therefore the claim was that the VW mean shapes display a highly significant difference.

A potential culprit for the CC shape-ADHD paradox could be the Kendall shape space dimensionality. Note that Kendall shape space $\Sigma_2^{50}$ has dimension 96, and the likelihood for two shape data sets pointing to a difference between the two means increases with the dimensionality (see Bai and Saranadasa (1996), Wei et al (2016)), which in case of shape analysis translates in an decrease in the p-value with the increase in the number of landmarks used.

Note that from the icons of the two sample VW-mean shapes, visually, it seems that the two populations have fairly close VW-means ( see Figure \ref{fig:means}).

\begin{figure}[H]%
    \centering
    \subfloat[VW mean of normal corpus ]{{\includegraphics[width=6cm]{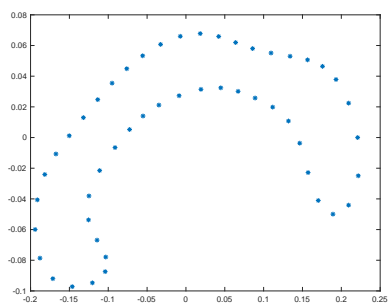} }}%
    \qquad
    \subfloat[VW mean of diagnosed ADHD corpus ]{{\includegraphics[width=6cm]{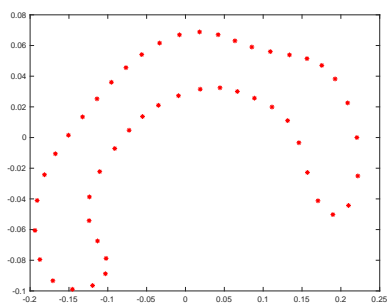} }}%
    \qquad
    \subfloat[VW means - normal (blue) vs ADHD (red) superimposed ]{\includegraphics[scale = 0.2]{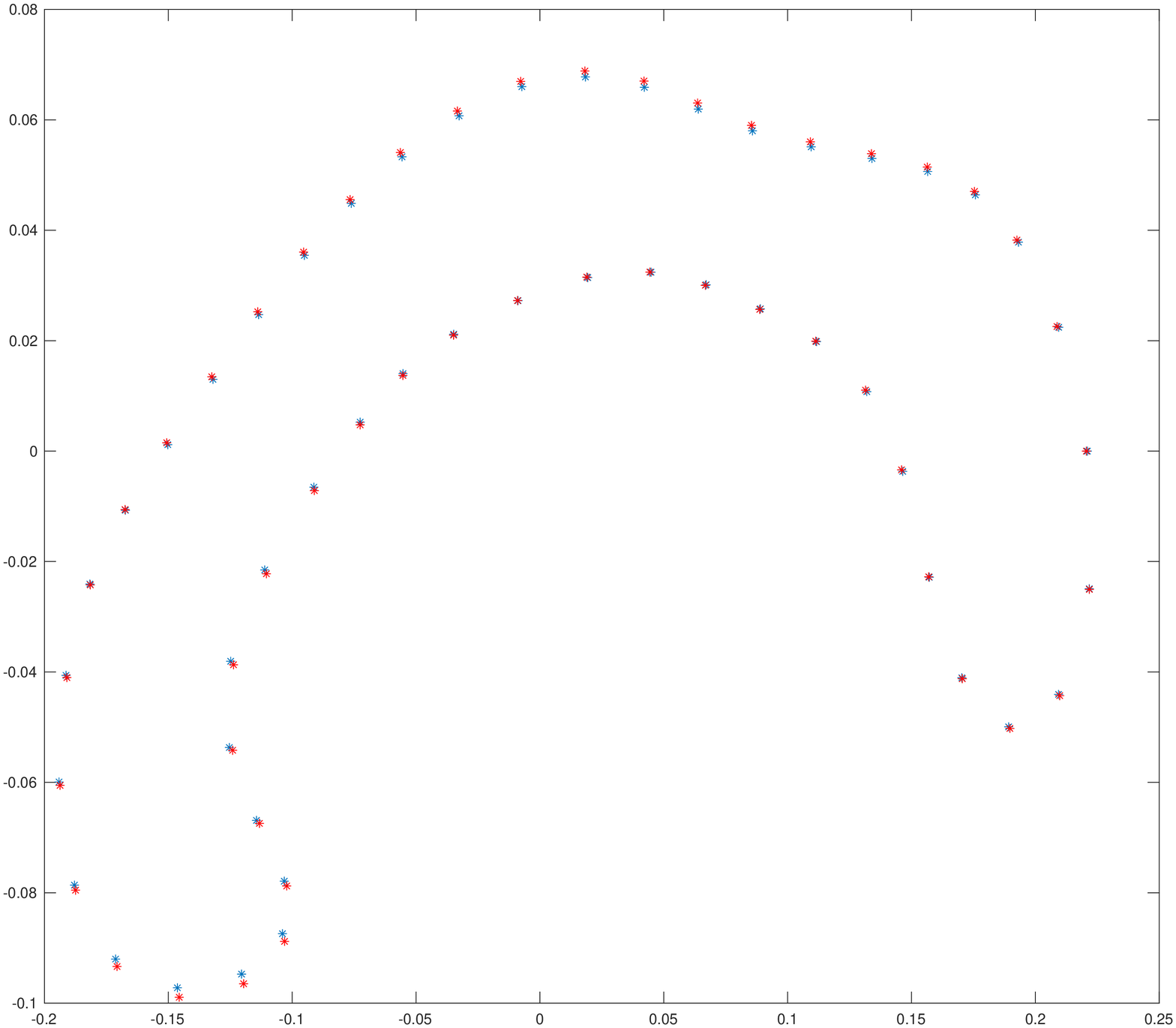}}%
    \caption{\small VW sample means: normal (upper-left), ADHD diagnosed (upper-right) and superimposed (lower)}%
    \label{fig:means}%
\end{figure}

\newpage
One the other side, one possibility for failing to reject the equality of the two shape distributions of the CC midsections of ADHD vs clinically normal youth, may be the inclusion in the CC data set of false ADHD positives. The incidence of ADHD seemed to have jumped by a high percentage in recent years, and there are skeptics in the Psychiatry community, about such a large increase (see Frances(2016)). In case of a high false ADHD detection rate, our analysis could confirm their suspicions.

\end{document}